
\magnification =\magstep1
\baselineskip 24pt
\hsize 33pc
\vsize 44pc
\rightline { IP/BBSR/91-51}
\rightline { December, 1991}
\vskip .5in

\def \clg {{\cal G}}

\centerline {\bf AN\hskip .1in ALGORITHM\hskip .1in TO\hskip .1in
GENERATE\hskip .1in CLASSICAL\hskip .1in SOLUTIONS}
\centerline {\bf OF\hskip .1in STRING\hskip .1in EFFECTIVE\hskip .1in ACTION}
\vskip .6in
\centerline {\bf Supriya K. Kar, S. Pratik Khastgir and Alok Kumar}
\vskip .1in
\centerline {\it Institute of Physics}
\centerline {\it Bhubaneswar, 751005}
\centerline {INDIA}
\vskip 1in
\centerline {\bf {$\cal ABSTRACT $}}
\vskip .4in
\par

It is shown explicitly, that a number of solutions for the background
field equations of the string effective action
in space-time dimension $D$ can be generated from any known lower
dimensional solution, when background fields have only time dependence.
An application of the result to the two dimensional charged black hole is
presented. The  case of background with more general coordinate
dependence is also discussed.

\vfil
\eject

Recently, there has been tremendous interest in the
search for nontrivial classical solutions $^1$ of the background field
equations of string theories. Solutions corresponding to
black holes $^{2,3,4,5,6}$, strings $^{7,8}$, branes $^9$ and
monopoles $^{10}$ have been reported in the literature. An interesting
aspect of these solutions is that, many of them can be shown to correspond
to exact conformal field theory. For example, black hole in two space-time
dimensions correspond to the coset $\; SL(2,R)/U(1)\; $ in conformal
field theory $^2$ and has recently been studied extensively by formulating
it as a gauged Wess-Zumino-Witten $(WZW)$ model. Two dimensional charged
black hole $^{5,6}$ as well as three dimensional charged black string
have also been formulated as gauged $WZW$ models $^{6,8}$.

The low energy string effective action was studied recently from a different
angle by Meissner and Veneziano $^{11,12}$. They showed that the
effective action in space-time dimension $\; D=d+1\; $ is invariant under an
$\; O(d,d)\; $ group of symmetry transformations when graviton, dilaton and
second rank antisymmetric tensor $B$ are the only nonvanishing background
fields and depend only on time coordinate. Low energy string effective
action, as well as the corresponding field equations have been
written down in manifestly $\; O(d,d)\; $ invariant form $^{12}$.
Therefore, given a solution of the background field equations, new
solutions can be obtained by $\; O(d,d)\; $ transformations.
Application of the $\; O(d,d)\;$ transformation in the context of two
dimensional charged black hole was done by two of the present authors in
Ref. 5. In Ref. 13, the idea of $\; O(d,d)\; $ invariance was
generalized to the case when background fields have more than just
time dependence. It was shown that, the string effective action has
an  $\; O({\tilde d},{\tilde d})\; $ invariance if the background is
independent of $ \tilde{ d} $ number of coordinates. In Ref. 14
interesting applications of $\; O(d,d)\; $ transformations were
presented and in Ref. ${15}$ the idea of
$\; O(d,d)\; $ invariance was generalized to heterotic strings.

In this paper, we study the field equations and effective action of Refs.11
and 13. First for only time dependent case of Ref. 11 , we show that,
a number of solutions of the graviton, dilaton and second rank antisymmetric
tensor background field equations, of the low energy string effective action,
can be generated  in space-time dimension '$D$' from a given classical
solution in any lower dimension. We apply these results to the two
dimensional charged black holes $^{5,6}$. We then discuss
the case when background fields have more than just time dependence.$^{13}$
The key to generate these solutions is a proper parametrization of the
 $D$- dimensional metric and antisymmetric tensor
 in terms of the lower dimensional ones.

Our starting point is the genus zero low energy effective action for closed
( super ) strings in the limit when string tension $\alpha '\rightarrow 0 $.
 Restricting to the graviton, dilaton and  antisymmetric tensor
field, this action in '$D$' space-time dimensions $\; D=d+1\; $ is written
as $^{11}$,
$$ S=\int d^{D}x \sqrt{-detG}\;
e^{-\phi}\big [V
-R^{(D)}-G^{\mu\nu}\partial_{\mu}\phi\partial_{\nu}\phi-{1\over
12}H_{\mu\nu\rho}H^{\mu\nu\rho}\big ]\eqno(1)$$

\noindent where  $\phi$ is the dilaton field, $G_{\mu\nu}$ is the $D$
-dimensional
 metric and $H_{\mu\nu\rho}$
is the field strength for the antisymmetric tensor field $B_{\mu\nu}$:
$$H_{\mu\nu\rho}=\partial_{\mu}B_{\nu\rho}+cyclic. \eqno(2)$$

\noindent $V$ in Eq.(1) contains cosmological constant as well as the dilaton
 potential. For many examples of interest $V$ is just a constant $^{5,12}$.

As in Refs. 11 and 12, we now investigate the solutions of the field equations
of the action (1) when $G$ and $B$ are functions of time only. In this case,
gauge symmetries of the action, Eq.(1), allow $G$ and $B$ to be always
 brought in the form $^{11}$:

$$G=\bigg (\matrix {{-1} & {0}\cr {0} &{{ \clg }(t)} \cr }\bigg )\quad ,
\qquad   B=\bigg (\matrix
{{0} & {0} \cr {0} & {{\cal B} (t)} \cr}\bigg ).\eqno(3)$$
where ${{\clg }(t)}$ and ${\cal B}(t)$ are $\; d\times d\; $
matrices. The action (1) for this case can be rewritten as $^{11}$,
$$ S=\int dt \sqrt{det\clg}\;
e^{-\phi}\big [V -2{{\partial }_{0}}^{2}(ln\sqrt {det \clg})-
({{\partial }_{0}}(ln\sqrt {det \clg}))
^{2}+ {1\over 4}Tr({{\partial }_{0}} \clg)({{\partial }_{0}} \clg^{-1})$$
$$+{({{\partial }_{0}} \phi )}^{2}+{1\over 4}
Tr({\clg }^{-1}({{\partial }_{0}}{\cal B})
{{\clg }^{-1}{({{\partial }_{0}}{\cal B})}})\big ].\eqno(4)$$

\noindent By redefining the
dilaton field : $$\Phi=\phi-ln\sqrt{det \clg}, \eqno(5)$$
one can write Eq.(4) as $^{11}$,
$$ S=\int dt\; e^{-\Phi}[V +{\dot {\Phi }}^{2}-{1\over 4}
Tr({\clg }^{-1} {\dot {\clg }})^{2}+{1\over 4}
Tr{{({\clg }^{-1} {\dot {\cal B }})}^{2}}].\eqno(6)$$

\noindent Equations of motion for these fields $^{11}$ are:
$$(\dot \Phi)^{2}-{1\over 4}Tr[({\clg }^{-1} \dot {\clg} )({\clg }^{-1}
\dot {\clg })]+ {1\over 4}Tr[({\clg }^{-1} \dot {\cal B} )({\clg }^{-1}
\dot {\cal B })]-V =0 ,\eqno(7)$$
$$({\dot {\Phi }})^{2}-2{\ddot {\Phi }}+{1\over 4}Tr[({{\clg }^{-1}
{\dot {\clg }}})({\clg }^{-1}{\dot {\clg }})]- {1\over 4}Tr[({{\clg }^{-1}
{\dot {\cal B }}})({\clg }^{-1}{\dot {\cal B }})]-V + {{\partial V}
\over{\partial {\Phi }}}=0\eqno(8)$$
$$-\dot {\Phi }{\dot {\clg }}+{\ddot {\clg }}-{\dot {\clg}}{\clg }^{-1}
{\dot {\clg }}- {\dot {\cal B}}{\clg }^{-1}{\dot {\cal B }}=0\eqno(9) $$
\noindent and
$$-\dot {\Phi }{\dot {\cal B }}+{\ddot {\cal B }}-{\dot {\cal B}}{\clg }^{-1}
{\dot {\clg }}- {\dot {\clg}}{\clg }^{-1}{\dot {\cal B }}=0 ,\eqno(10) $$

\noindent where Eqs.(8), (9) and (10) follow from the variations with respect
 to the fields $\Phi$, ${\clg }_{ij}$  and ${\cal B}_{ij}$ of action (6).
Equation (7), which is obtained directly from the variation of the action (1),
with respect to $G_{00}$, is also called the "zero energy condition"$^{11}$. It
has also been pointed out in Ref. 11, that Eqs.(7)-(10) are the
complete set of equations of motion for this set of fields. In Ref. 16, these
equations were solved for arbitrary $D$ when $\clg$ is diagonal and
${\cal B}$ is absent. For $D=2$, the solution of Ref. 16 is identical to the
 uncharged black hole solution $^{2,3}$, when roles of space and time are
interchanged. In Ref.5, two of us obtained a solution of Eqs. (7)-(10) in the
presence of off-diagonal terms in the metric $\clg$ and vanishing ${\cal B}$,
for $D=3$. There  we showed that, this solution can be
interpreted as two dimensional charged black hole solution upon
compactification of one of the space dimensions and by interchanging
the roles of space and time as in Ref. 16.

Now we show that, a number of solutions of Eqs.(7)-(10) for $d$-dimensional
metrices $\clg $ and ${\cal B}$ can be obtained from a given solution
of similar equations for $\hat{ d} \; (\equiv d-p)$-dimensional metrices
$\hat G$ and ${\hat {\cal B}}$. To obtain solutions for $\clg $ and
${\cal B}$ from $\hat G$ and $\hat {\cal B}$, we parametrize the
$\; d\times d\; $ matrices $\clg $  and ${\cal B}$ as,
$$\clg\; =\;\Bigg (\matrix {{{\hat G}(t)} & {{1\over2}{\hat G}(t)\; b^T}
\cr {{1\over2}b\; {\hat G}(t)} & {\psi +{1\over 4}b\; {\hat G}(t)\; b^T}
\cr }\Bigg )\; ,\;\;
{\cal B}\; =\;\Bigg (\matrix {{{\hat {\cal B}}(t)} & {{1\over2}
{\hat {\cal B}}(t)\; b^T}
\cr {{1\over2}b\; {\hat {\cal B}(t)}} & {{1\over 4}b\; {\hat {\cal B}}(t)\;
b^T}
\cr }\Bigg ),\eqno(11)$$

\noindent where ${\hat G}(t)$ and ${\hat {\cal B}}(t)$ are
$\;\hat d\times\hat d\; $ matrices, $\psi $ and $b$ are respectively
$\; p\times p\; $ and $\; p\times \hat d\; $ matrices and $\psi $ is
nonsingular. Constant metric ${\clg}$ of the form in Eq.(11) has been used
previously for string compactifications $^{17}$. For our case however, we note
that, all elements of $\clg$ and ${\cal B}$ in Eqs.(11) have nontrivial time
dependence. We now make a simplifying assumption that $\psi $ and $b$ are
constant matrices. From Eq.(11) one obtains,
$${\clg}^{-1}\; =\;\Bigg (\matrix {{{\hat G}^{-1}+{1\over4}b^T{\psi }^{-1}b}
& {-{1\over2}b^T\; {\psi }^{-1}}\cr {-{1\over2}{\psi }^{-1}\; b} &
{{\psi }^{-1}} \cr }\Bigg )\eqno(12)$$

\noindent and $$det {\clg }\; =\; (det{\hat G})\; (det{\psi }).\eqno(13)$$

\noindent Since we have assumed that $\psi $ and $b$ are constant matrices,
we have from Eqs.(11) and (12):
$$\dot {\clg }\; =\; \Bigg (\matrix {{\dot {\hat G}} &
{{1\over2}{\dot {\hat G}}\; b^T} \cr {{1\over2}b\; {\dot {\hat G}}} &
{{1\over4}b\; {\dot {\hat G}}\; b^T}\cr }\Bigg )\; ,
\;\;\dot {\cal B }\; =\; \Bigg (\matrix {{\dot {\hat {\cal B}}} &
{{1\over2}{\dot {\hat {\cal B}}}\; b^T} \cr {{1\over2}b\; {\dot {\hat
{\cal B}}}} & {{1\over4}b\; {\dot {\hat {\cal B}}}\; b^T}\cr }\Bigg ) ,
\eqno(14)$$

$$\ddot {\clg }\; =\; \Bigg (\matrix {{\ddot {\hat G}} & {{1\over2}{\ddot
{\hat G}}\; b^T}\cr {{1\over2}b\; {\ddot {\hat G}}} & {{1\over4}b\; {\ddot
{\hat G}\; b^T}}\cr }\Bigg )\; ,\;\;
\ddot {\cal B }\; =\; \Bigg (\matrix {{\ddot {\hat {\cal B}}} &
{{1\over2}{\ddot {\hat {\cal B}}}\; b^T}\cr {{1\over2}b\; {\ddot {\hat
{\cal B}}}} & {{1\over4}b\; {\ddot
{\hat {\cal B}}\; b^T}}\cr }\Bigg )\eqno(15)$$

\noindent and
$$\dot {({\clg }^{-1})}\; =\; \Bigg (\matrix {{\dot {({\hat G}^{-1})}} & {0}\cr
{0} & {0}\cr }\Bigg ).\eqno(16)$$

\noindent We now use equations (11)-(16) to simplify the background field
equations (7)-(10). Using (12) and (14) one can show that,
$$Tr\big [\big ({\clg }^{-1}{\dot {\clg }} \big )\big ({\clg }^{-1} {\dot
{\clg }}\big )\big ]\; =\; Tr\big [\big ({\hat G}^{-1}{\dot {\hat
G}}\big )\big ({\hat G}^{-1}{\dot {\hat G}}\big )\big ]\eqno(17)$$
\noindent and
$$Tr\big [\big ({\clg }^{-1}{\dot {\cal B}} \big )\big ({\clg }^{-1} {\dot
{\cal B}}\big )\big ]\; =\; Tr\big [\big ({\hat G}^{-1}{\dot {\hat
{\cal B}}}\big )\big ({\hat G}^{-1}{\dot {\hat {\cal B}}}\big )\big ] .
\eqno(18)$$

\noindent Therefore, Eq.(7) can be rewritten as,
$$(\dot \Phi)^{2}-{1\over 4}Tr[({\hat G}^{-1} \dot {\hat G} )({\hat G}^{-1}
\dot {\hat G} )]+{1\over 4}Tr[({\hat G}^{-1} \dot {\hat {\cal B}} )
({\hat G}^{-1}\dot {\hat {\cal B}})]-V =0. \eqno(19)$$

\noindent Similarly, Eq.(8) can be rewritten as
$${(\dot {\Phi })}^{2}-2\ddot {\Phi }+{1\over 4}Tr[({\hat G}^{-1}
\dot {\hat G} )({\hat G}^{-1}\dot {\hat G} )]-{1\over 4}Tr[({\hat G}^{-1}
\dot {\hat {\cal B}})({\hat G}^{-1}\dot {\hat {\cal B}} )]-V +
{{\partial V}\over {\partial {\Phi }}}=0 .\eqno(20)$$

\noindent Also, using Eqs.(12), (14) and (15), Eq. (9) can be rewritten as a
matrix equation:
$$-\dot {\Phi }\;\Bigg (\matrix {{\dot {\hat G}} &
{{1\over2}{\dot {\hat G}}\; b^T} \cr {{1\over2}b\; {\dot {\hat G}}} &
{{1\over4}b\; {\dot {\hat G}}\; b^T}\cr }\Bigg )\; +
\Bigg (\matrix {{\ddot {\hat G}} & {{1\over2}{\ddot
{\hat G}}\; b^T}\cr {{1\over2}b\; {\ddot {\hat G}}} & {{1\over4}b\; {\ddot
{\hat G}\; b^T}}\cr }\Bigg )$$
$$\; -\; \Bigg (\matrix {{\dot {\hat G}\; {\hat G}^{-1}\dot {\hat G}} &
{{1\over2}(\dot {\hat G}\; {\hat G}^{-1}\dot {\hat G})\; b^T} \cr
{{1\over2}b\; (\dot {\hat G}\; {\hat G}^{-1}\dot {\hat G})}
& {{1\over4} b\; (\dot {\hat G}\; {\hat G}^{-1}\; \dot {\hat G})\; b^T}\cr }
\Bigg )\; -\; \Bigg (\matrix {{\dot {\hat {\cal B}}\; {\hat G}^{-1}
\dot {\hat {\cal B}}} & {{1\over2}(\dot {\hat {\cal B}}\; {\hat G}^{-1}
\dot {\hat {\cal B}})\; b^T} \cr {{1\over2}b\; (\dot {\hat {\cal B}}\;
{\hat G}^{-1}\dot {\hat {\cal B}})} & {{1\over4} b\; (\dot {\hat
{\cal B}}\; {\hat G}^{-1}\; \dot {\hat {\cal B}})\; b^T}\cr }
\Bigg ) =\; 0\eqno(21)$$

\noindent Equation (21) gives rise to four matrix-equations. However, it is
recognized that all of them are same and can be written as a single matrix
equation:
$$-\dot {\Phi }\dot {\hat G}\; +\; \ddot {\hat G}\; -\;\dot {\hat
G}\; {{\hat G}^{-1}}\dot {\hat G}\; -\;\dot {\hat {\cal B}}\;{{\hat G}^{-1}}
\dot {\hat {\cal B}}\; =\; 0.\eqno(22)$$

\noindent Eq.(10) can similarly be simplified to
$$-\dot {\Phi }\dot {\hat {\cal B}}\; +\; \ddot {\hat {\cal B}}\; -\;\dot {\hat
{\cal B}}\; {{\hat G}^{-1}}\dot {\hat G}\; -\;\dot {\hat {\clg }}\;
{{\hat G}^{-1}}\dot {\hat {\cal B}}\; =\; 0.\eqno(23)$$

\noindent We now use Eqs.(5) and (13) to write
$$\Phi\; =\;\hat\phi\; -\; ln\; {\sqrt {det\; \hat G}}\eqno(24)$$
\noindent where
$$\hat\phi\; =\;\big (\phi\; -\; ln\; {\sqrt {det\;\psi }}\big ).\eqno(25)$$
\noindent Therefore we note that, by making a constant shift from
${\phi }\;\rightarrow\; \hat{\phi} \; ,\;\Phi $ can also be thought of as the
"redefined" dilaton in $\hat d$-dimensions. Then we recognize that
Eqs.(19), (20), (22) and (23) are in fact the background field equations
in $\hat d$-dimensions. Our Eqs.(11) and (25) therefore provide a
prescription for generating solution of the field equations for the
$d$-dimensional fields $\clg $ and ${\cal B}$ from the $\hat d$-dimensional
ones, $i.e$. $\hat G$, $\hat {\cal B}$.

As an application of our result, we now consider the case of
two dimensional charged black hole. We start with $\hat d=1$ so that
$\hat G$, ${\hat {\cal B}}$ are single functions and a solution for Eqs.(19),
(20), (22) and (23) is given by $^{16}$,
$$\hat G\;\equiv\;\hat g\; tanh^2\; t\; ,\;\; {\hat {\cal B}}=0\eqno(26)$$
\noindent and
$$\hat\phi\;\equiv\; -\; ln\; ( cosh^2\; t)\; +\; c_1,\eqno(27)$$

\noindent where $\hat g$ and $c_1$ are constants.
Then, using Eq.(11), one can generate $(p+1)$-dimensional solutions of the
type,
$$\clg\; =\; \Bigg (\matrix {{{\hat g}\; tanh^2\; t} & {{1\over2}b_i\;
\hat g\; tan\; h^2\; t}\cr  {{1\over2}b_j\; \hat g\; tan\; h^2\; t} &
{{\psi }_{ij}+{1\over4}\hat g\; b_i\; b_j\; tanh^2\; t}\cr }\Bigg )\;
,\;\; {\cal B}=0.\eqno(28)$$
\noindent where $(i,j=1,\; 2,\; . . . . p)$. Also from (25),
$$\phi\; =\; -\; ln\;\big (cos\; h^2\; t\big )\; +\; c_2 ,\eqno(29)$$
\noindent where $c_2$ is a constant different from $c_1$.
For $b_{i}=0$, $i>1$ and $b_{1}=b$ solution (28) for $\clg$
implies that metric G in Eq.(3) is of the form:
$$G\; =\; \Bigg (\matrix {{-1} & {0} & {0}\cr {\; 0} & {\hat g\; tan\;
h^2\; t} & {{1\over2}b\;\hat g\; tan\; h^2\; t}\cr {\; 0} &
{{1\over2}b\;\hat g \; tan\; h^ 2\; t} & {{\psi }_{11}+
{1\over4}{b^2}\; {\hat g}\; {tan\; h^2\; t}}\cr }\Bigg ).\eqno(30)$$

\noindent Now, if third dimension is treated as a compact one, then G
in Eq.(30) gives rise, in two space-time dimensions, to the metric
${\tilde G}_{\mu\nu}$, gauge field ${\tilde A}_{\mu}$ and Higgs field
$\tilde\psi $ as $^{5}$
$${\tilde G}_{\mu\nu}\equiv\bigg (\matrix {{-1} & {0}\cr {0} & {\hat g\;
tanh^2\; t}\cr }\bigg ),\eqno(31)$$
$${\tilde A}_{\mu}\equiv\big (0\; ,\; b\; \hat g\; tanh^2\; t\big )
\eqno(32)$$
\noindent and
$$\tilde\psi\;\equiv {\psi }_{11}+{1\over 4}b^2\hat g\; tanh^2\; t .\eqno(33)$$

The solutions in Eqs.(31)-(33) can be interpreted as the
two dimensional charged black hole solution $^{5,6}$ by changing the
roles of space and time.  In Ref. 5 it was also shown that the solutions
Eqs.(31)-(33), correspond to the coset of the
type $\; [SL(2,R)\times U(1)]/U(1)\; $ in the conformal field theory,
which is formulated as the gauged $WZW$ model $^6$. In general, for
${b_i\neq 0}$, one has ${\tilde G}_{\mu\nu}$ as in Eq.(31)
together with $[U(1)]^p$ gauged fields,
$$\tilde {A^i_{\mu}}\equiv\big (0\; ,\; b_i\;\hat g\;tanh^2\; t\big ) ,
\eqno(34)$$
\noindent and Higgs fields,
$${\tilde\psi }_{ij}\; =\; \big ({\psi }_{ij} + {1\over4}\hat g\; b_i
\; b_j\; tanh^2\; t\big ).\eqno(35)$$

\noindent By comparing with the gauged $WZW$ model, it can be shown that the
solution of the type given in
Eqs.(29), (31), (34) and (35) corresponds to the coset
$\; [SL(2,R)\times [U(1)]^p/U(1)]\; $ in conformal field theory.

We now discuss the generalization of these results to the case when background
has more general coordinate dependence, but is independent of at
least one of them. Following the second paper of
Ref. 13, we split the space- time coordinates
$x^{\mu }$ in Eq.(1) into two sets $y^m$ and ${\tilde y}^{\alpha }$
$(1\leq m\leq {\tilde d}$, $1\leq \alpha\leq {D-{\tilde d}}$) and consider
backgrounds independent of $y^m$. As in Ref. 13, we restrict to the
backgrounds: $G_{m\alpha }=0$, $B_{m\alpha }=0$. Then
background fields $G_{\mu\nu }$, $B_{\mu\nu }$ can be written in the
form,

$$G_{\mu\nu }=\bigg
(\matrix {{{\tilde {\clg }}_{\alpha\beta }({\tilde y})} & {0}\cr {0}
& {{\clg }_{mn}({\tilde y})}\cr }\bigg )\; ,\;\;
B_{\mu\nu }=\bigg (\matrix { {{\tilde {\cal B}}_{\alpha\beta }{({\tilde y})}}
& {0}\cr {0} &
{{\cal B}_{mn}({\tilde y})}\cr }\bigg ).\eqno(36)$$
\noindent For this case, after an integration by parts, action (1)
can be written as $^{13}$:
$$S\; =\; \int\; d^{\tilde d}{y}\;\int\; d^{(D-{\tilde d})}{\tilde y}\;
\sqrt {- det {{\tilde \clg }}({\tilde y})}
\; e^{-\Phi }\big [V-{{\tilde \clg }}^{\alpha\beta }\; {\tilde {\partial }}_
{\alpha }{\Phi }\; {\tilde {\partial }}_{\beta }{\Phi }- {1\over4}
{{\tilde \clg }}^{\alpha\beta }Tr\big ({\tilde {\partial }}_{\alpha }{\clg }\;
{\tilde {\partial }}_{\beta }{{\clg }^{-1}}\big )\;$$
$$ -\; {1\over4}{\tilde {\clg }}^{\alpha\beta }Tr\big
({{\clg }^{-1}}{\tilde {\partial }}_{\alpha }{\cal B}\; {{\clg }^{-1}}
{\tilde {\partial }}_{\beta }{\cal B}\big )\; -\; R^{({D-{\tilde d}})}
{({\tilde {\clg }})}\; -\;
{1\over{12}}{\tilde {\cal H}}_{\alpha\beta\gamma }{\tilde
{\cal H}}^{\alpha\beta\gamma }\big ]\eqno(37)$$

\noindent where $\Phi =\phi -ln\;\sqrt {det\; {{\clg }({\tilde y})}}$.
We now do a similar substitution as in Eq.(11), $ie$.

$$\clg\; =\;\Bigg (\matrix {{{\hat G}({\tilde y})} &
{{1\over2}{\hat G}({\tilde y})\; b^T}
\cr {{1\over2}b\; {\hat G}({\tilde y})} &
{\psi +{1\over 4}b\; {\hat G}({\tilde y})\; b^T}
\cr }\Bigg )\; ,\;\;
{\cal B}\; =\;\Bigg (\matrix {{{\hat {\cal B}}({\tilde y})} & {{1\over2}
{\hat {\cal B}}({\tilde y})\; b^T}
\cr {{1\over2}b\; {\hat {\cal B}({\tilde y})}} &
{{1\over 4}b\; {\hat {\cal B}}({\tilde y})\;
b^T} \cr }\Bigg ), \eqno(38)$$

\noindent where ${{\hat G}({\tilde y})}$ and
${{\hat B}({\tilde y})}$ are $\;\hat
d\times\hat d\; $ matrices and  $\psi $, $b$ are as defined earlier but
$\hat d $ is now such that,
${\hat d}={\tilde d}-p$.
We now use Eq.(38) to rewrite the action (37) as,

$$S\; =\; \int\; d^{\tilde d}y\;\int\; d^{(D-{\tilde
d})}{\tilde y}\;\sqrt {- det {\tilde {\clg }}({\tilde y})}
\; e^{-\Phi }\big [V-{\tilde {\clg }}^{\alpha\beta }\; {\tilde {\partial }}_
{\alpha }{\Phi }\; {\tilde {\partial }}_{\beta }{\Phi }
\; -\; {1\over4}{\tilde {\clg }}^{\alpha\beta }Tr\big
({\tilde {\partial }}_{\alpha }{\hat G}\; {\tilde {\partial }}_{\beta }
{{\hat G}^{-1}}\big )$$
$$\; -\; {1\over4}{\tilde {\clg }}^{\alpha\beta }Tr\big
({{\hat G }^{-1}}{\tilde {\partial }}_{\alpha }{\hat {\cal B}}\; {{\hat
G}^{-1}}
{\tilde {\partial }}_{\beta }{\hat {\cal B}}\big )\; -\; R^{(D-{\tilde d})}
{({\tilde {\cal G}})}\; -\;{1\over{12}}{\tilde {\cal H}}_{\alpha\beta\gamma }
{\tilde {\cal H}}^{\alpha\beta\gamma }\big ]\eqno(39)$$

We see that the action (37) has retained its form after the substitution
(38) and therefore
Eqs. of motion will also preserve their form. Hence, using Eq.(38),
new higher dimensional solutions ${\clg }$ and ${\cal B}$ can be
generated from the known lower dimensional ones, $i.e.$ $\hat{G}$ and
$\hat {\cal B}$ as before. However, unlike the case of only time
dependent background, this has been demonstrated for metric and
antisymmetric tensor of the block diagonal form only, $i.e.$
$G_{m\alpha } =\; B_{m\alpha } =\; 0$

To conclude, we have shown that a number of
solutions (classified by constant parameters $b_{ia}$ and ${\psi }_{ij}$),
of the background field equations can be obtained
in higher dimension from a given solution in any lower dimension.
It will be interesting to generalize the
results to heterotic string $^{15}$.

\vfil
\eject

\noindent {\bf References:}

\item {[1]} G. W. Gibbons and K. Maeda, Nucl. Phys. {\bf B298}, 741 (1988);
C. G. Callan, R. C. Myers and M. J. Perry, Nucl. Phys. {\bf B311},673
(1988/89); H. J. Vega and N. Sanchez, Nucl. Phys. {\bf B309}, 552
(1988); 557 (1988).

\item {[2]} E. Witten, Phys. Rev. {\bf D44}, 314 (1991);
R. Dijgraaf, E. Verlinde and H. Verlinde, Institute for Advanced Study
preprint IASSNS-HEP-91/22 (1991).

\item {[3]} G. Mandal, A. M. Sengupta and S. Wadia, Mod. Phys. Lett.
{\bf A6}, 1685 (1991);
S. P. de Alwis and J. Lykken, Phys. Lett. {\bf B269}, 264 (1991);
A. Tseytlin, John Hopkins preprint JHU-TIPAC-91009; A. Giveon, LBL
preprint LBL-30671.

\item {[4]} D. Garfinkle, G. T. Horowitz and A. Strominger,
Phys. Rev. {\bf D43}, 3140 (1991).

\item {[5]} S. P. Khastgir and A. Kumar, Mod. Phys. Lett. {\bf A6},
3365 (1991).

\item {[6]} N. Ishibashi, M. Li and A. R. Steif, Phys. Rev. Lett.
{\bf67}, 3336 (1991).

\item {[7]} G. T. Horowitz and A. Strominger, Nucl. Phys. {\bf B360},
197 (1991);
J. H. Horne, G. T. Horowitz and A. R. Steif, Univ. of California,
Santa Barbara preprint UCSBTH-91-53, October 1991.

\item {[8]} J. H. Horne and G. T. Horowitz, Univ. of California,
Santa Barbara preprint UCSBTH-91-39, July 1991.

\item {[9]} S. B. Giddings and A. Strominger, Phys. Rev. Lett. {\bf67},
2930 (1991).

\item {[10]} J. A. Harvey and J. Liu, Phys. Lett. {\bf B268}, 40 (1991).

\item {[11]} K. A. Meissner and G. Veneziano, Phys. Lett. {\bf B267},
33 (1991).

\item {[12]} K. A. Meissner and G. Veneziano, Mod. Phys. Lett. {\bf A6},
3397 (1991).

\item {[13]} A. Sen, Phys. Lett. {\bf B271}, 295 (1991);
A. Sen, Tata preprint, TIFR/TH/91-37, August 1991.

\item {[14]} M. Gasperini, J. Maharana and G. Veneziano, CERN Theory
preprint CERN-TH-6214/91 (1991); M. Gasperini and G.Veneziano, CERN
Theory preprint CERN-TH-6321/91 (1991).

\item {[15]} S. F. Hassan and A. Sen, Tata preprint, TIFR/TH/91-40,
September 1991.

\item {[16]} G. Veneziano, Phys. Lett. {\bf B265}, 287 (1991).

\item {[17]} A. Giveon, E. Rabinovici and G. Veneziano, Nucl. Phys.
{\bf B322}, 167 (1989).

\end